\documentclass[%
 reprint,twocolumn,
superscriptaddress,
nofootinbib,
 pre,amsmath,amssymb,
]{revtex4-2}

\usepackage{url}
\usepackage{graphicx}
\usepackage{bm}
\usepackage[dvipsnames]{xcolor}
\usepackage{multirow}

\begin{document}

\newtheorem{theorem}{Theorem}
\newtheorem{definition}{Definition}
\newtheorem{lemma}{Lemma}
\newtheorem{proposition}{Proposition}
\newtheorem{remark}{Remark}
\newtheorem{corollary}{Corollary}
\newtheorem{example}{Example}

\newcommand{\eref}[1]{Eq.~(\ref{#1})}%
\def\bea{\begin{eqnarray}}
\def\eea{\end{eqnarray}}

\newcommand{\Eref}[1]{Equation~(\ref{#1})}%
\newcommand{\fref}[1]{Fig.~\ref{#1}} %
\newcommand{\Fref}[1]{Figure~\ref{#1}}%
\newcommand{\sref}[1]{Sec.~\ref{#1}}%
\newcommand{\Sref}[1]{Section~\ref{#1}}%
\newcommand{\aref}[1]{Appendix~\ref{#1}}%
\newcommand{\sgn}[1]{\mathrm{sgn}({#1})}%
\newcommand{\erfc}{\mathrm{erfc}}%
\newcommand{\erf}{\mathrm{erf}}%


\title{Geometric Brownian Motion under Stochastic Resetting: A Stationary yet Non-ergodic Process}

\author{Viktor Stojkoski}
\email{vstojkoski@manu.edu.mk}
\affiliation{Faculty of Economics, Ss.~Cyril and Methodius University, 1000 Skopje, Macedonia}
\affiliation{Research Center for Computer Science and Information Technologies, Macedonian Academy of Sciences and Arts, Bul. Krste Misirkov 2, 1000 Skopje, Macedonia}
 
\author{Trifce Sandev}
\email{trifce.sandev@manu.edu.mk}
\affiliation{Research Center for Computer Science and Information Technologies, Macedonian Academy of Sciences and Arts, Bul. Krste Misirkov 2, 1000 Skopje, Macedonia}
\affiliation{Institute of Physics \& Astronomy, University of Potsdam, D-14776 Potsdam-Golm, Germany}
\affiliation{Institute of Physics, Faculty of Natural Sciences and Mathematics, Ss.~Cyril and Methodius University, Arhimedova 3, 1000 Skopje, Macedonia}

\author{Ljupco Kocarev}
\email{lkocarev@manu.edu.mk}
\affiliation{Research Center for Computer Science and Information Technologies, Macedonian Academy of Sciences and Arts, Bul. Krste Misirkov 2, 1000 Skopje, Macedonia} \affiliation{Faculty of Computer Science and Engineering, Ss.~Cyril and Methodius University, P.O. Box 393, 1000 Skopje, Macedonia}

\author{Arnab Pal}%
\thanks{Corresponding author}
 \email{arnabpal@mail.tau.ac.il}
\affiliation{School of Chemistry, The Center for Physics and Chemistry of Living Systems, Tel Aviv University, Tel Aviv 6997801, Israel}%




\date{\today}

\begin{abstract}
We study the effects of stochastic resetting on geometric Brownian motion with drift (GBM), a canonical stochastic multiplicative process for non-stationary and non-ergodic dynamics. Resetting is a sudden interruption of a process, which consecutively renews its dynamics. We show that, although resetting renders GBM stationary, the resulting process remains non-ergodic. 
Quite surprisingly, the effect of resetting is pivotal in manifesting the non-ergodic behavior. In particular, we observe three different long-time regimes: a quenched state, an unstable and a stable annealed state depending on the resetting strength. Notably, in the last regime, the system is self-averaging and thus the sample average will always mimic ergodic behavior establishing a stand alone feature for GBM under resetting.  Crucially, the above-mentioned regimes are well separated by a self-averaging time period which can be minimized by an optimal resetting rate. Our results can be useful to interpret data emanating from stock market collapse or reconstitution of investment portfolios.

\end{abstract}

\maketitle


\section{Introduction}

Geometric Brownian motion (GBM) is a universal model for self-reproducing phenomena, such as population and wealth~\cite{braumann1983population}. Perhaps the
best-known application of GBM is in mathematical finance
(
the Black-Scholes model)
for asset pricing
\cite{black1973pricing,shah1997black}. GBM has also been used to model a myriad of other natural phenomena such as bacterial cell division, inheritance of fruit and flower size, body-mass distribution, rainfall, fragment sizes in rock crushing processes, etc. (see \cite{limpert2001log,aitchison1957lognormal} for a review).

Stochastic processes governed by GBM show unconstrained growth phenomena, thus they are non-ergodic and non-stationary \cite{peters2013ergodicity,cherstvy2017time}. Nonetheless, a prevalent real world observation conforms that self-reproduction is characterized by a stationary distribution that has power law tails, which hinders the practical implementation of the model~\cite{zipf2016human}. A natural way to \textit{invoke stationarity} is to adapt GBM with a stochastic resetting mechanism which intermittently stops the current dynamics only to restart again and has spurred extensive research interests recently in statistical physics \cite{evans2011diffusion,evans2011diffusionJPA,evans2020stochastic,majumdar2015dynamical,pal2015diffusion,pal2016diffusion,nagar2016diffusion,basu2019symmetric,singh2020resetting,gupta2020work,gupta2020stochastic,mendez2016characterization,gupta2014fluctuating,evans2013optimal,magoni2020ising,durang2014statistical,pal2017integral,ray2020space}, stochastic processes \cite{manrubia1999stochastic,zanette2020fat,kusmierz2014first,pal2017first,meylahn2015large,bodrova2019scaled,pal2019landau,pal2019firstV,chechkin2018random,kusmierz2019subdiffusive,belan2018restart,pal2019firstbranch,de2020optimization,domazetoski2020stochastic,boyer2017long,singh2021extremal} and in single particle experiments \cite{tal2020experimental,besga2020optimal}.
Furthermore, many natural phenomena described by GBM, often undergo catastrophes (reminiscent of resetting \cite{taleb2007black,brockwell1985extinction,dharmaraja2015continuous,di2003m,di2012double}) thus describing situations such as pandemics or sudden stock market crashes. Although these observations are intriguing, there is no detailed statistical analysis of GBM subject to stochastic resetting (srGBM) with a focus on long time statistics of self-reproducing resources or their ergodic properties where the latter is quite fundamental to various disciplines, ranging from economics to evolutionary biology \cite{peters2019ergodicity,stojkoski2019cooperation}. 
Only ensemble average properties of models somewhat similar to srGBM have been investigated in a handful of economics literature \cite{nirei2004income,guvenen2007learning,aoki2017zipf,gabaix2016dynamics,kou2002jump}, but nothing is known on the time-averaging. This letter exactly delves deeper into these central aspects. 

For brevity, the results are briefly summarized in the following.  We show that srGBM reaches a stationary state in the long time limit yet the process remains non-ergodic. Non-ergodicity in srGBM is realized in the long-time behavior of an average over a \textit{finite} sample of trajectories with three emerging regimes: i) a frozen state regime, ii) an unstable annealed regime, and iii) a stable annealed regime. The long time and short time behavior of the system in these regimes are separated by a critical time scale which depends strongly on the resetting rate. 
In the first regime, named after an analogy to the celebrated ``Random Energy Model'' by Derrida \cite{derrida1981random}, the long time behavior of the system is the same as in the standard GBM. This implies that resetting does not affect the non-ergodicity of the process. However, in the other two regimes, resetting non-trivially ramifies the self-averaging behavior 
leading to either an unstable or a stable 
long-time sample average. Importantly, in the last regime, for a large enough sample size, the system may always be self-averaging and thus the sample average will forever mimic ergodic behavior. Besides the emphasized effect on the ergodicity of the process, we also show that when resetting is Poissonian, there exists an optimal resetting rate that minimizes the critical self-averaging time thus displaying another intriguing feature of this study.

The paper is structured as follows. In \sref{model}, we introduce the model, provide preliminary results and discuss the simulation procedure. We present exact results for the moments and the probability density function for the GBM with stochastic resetting in \sref{GPM}. \sref{ergodic} and \sref{SAP} respectively are dedicated to discuss ergodic and self-averaging properties of GBM with stochastic resetting. We conclude in \sref{conclusions} with a summary of our work and future directions.

\section{Model} 
\label{model}

\subsection{Preliminaries}

Motion of a particle governed by srGBM is described by the following Langevin equation
\begin{align}
d x(t) &=(1-Z_{t}) x(t) \left[ \mu dt+ \sigma dW \right]+Z_{t} \left( x_0-x(t) \right),
\label{eq:srgbm-microscopic}
\end{align}
where $x(t)$ is the position of the particle (but could be self-reproducing resources such as biomass or capital) at time $t$, $dt$ denotes the infinitesimal time increment and
$dW$ is an infinitesimal Wiener increment, which is a normal variate with $\langle dW_t \rangle=0$ and $\langle dW_t dW_s \rangle =\delta(t-s)dt$. Here, $\mu$ and $\sigma$ are called the drift and noise amplitude. Resetting is introduced with a random variable $Z_{t}$ which takes the value $1$ when there is a resetting event in the time interval between $t$ and $t +dt$; otherwise, it is zero. Without any loss of generality, we also assume that resetting brings the particle back to its initial condition $x(0)=x_0$.

The solution to Eq.~\eqref{eq:srgbm-microscopic} can be found by interpreting srGBM as a renewal process: each resetting event renews the process at $x_0$ and between two such consecutive renewal events, the particle undergoes the simple GBM. Thus, between time points $0$ and $t$, only the last resetting event, occurring at the point 
\begin{align}
    t_{l}(t) = \max_{k \in \left[0,t\right]} k: \{ Z_{k} = 1 \},
    \label{eq:srgbm-solution-1-0}
\end{align}
is relevant and the solution to Eq.~\eqref{eq:srgbm-microscopic} reads (following Itô interpretation)
\begin{align}
      x(t) &= x_0~ e^{ (\mu - \frac{\sigma^2}{2}) \left[ t - t_{l}(t) \right] + \sigma \left[ W(t) - W(t_{l}(t)\right]}.
      \label{eq:srgbm-solution-2-0}
\end{align}
In what follows we will assume stochastic resetting so that the probability for a reset event is given by $P(Z_{t} = 1) = rdt$. In the limit when $dt \to 0$, this corresponds to an exponential resetting time density $f_r(t)=re^{-rt}$, and $t_l$ is distributed according to
\begin{align}
f(t_{l}|t)=\delta(t_{l}) e^{-rt}+re^{-r(t-t_{l})},
\label{last-time-pdf}
\end{align}
such that $\int_0^t ~dt_l f(t_l|t)=1$.
Intuitively, the first term on the RHS corresponds to the scenario when there is no resetting event up to time $t$ while the second one accounts for multiple resetting events.  Notably, writing stochastic solutions (such as \eref{eq:srgbm-solution-2-0}) on a single trajectory level in the presence of resetting is quite useful, as will be seen below. 
We further stress that \eref{eq:srgbm-solution-2-0} also holds for complex restart time distributions with a straightforward generalization of \eref{last-time-pdf} that can be obtained from Refs. \cite{pal2016diffusion,chechkin2018random}.

\begin{figure}[b!]
\includegraphics[width=8.6cm]{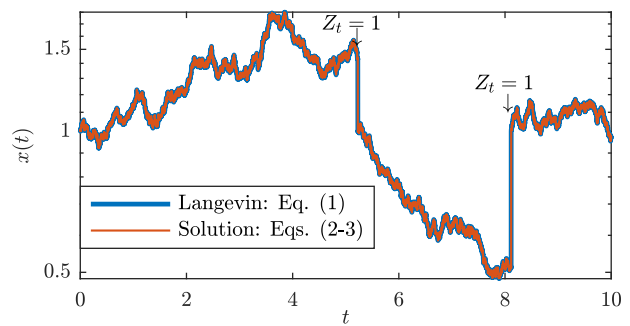} 
\caption{\textbf{srGBM dynamics}. Position of the particle evolves multiplicatively via \eref{eq:srgbm-microscopic} until a random event characterized by $Z_t = 1$ occurs. At this moment, position is reset to $x_0=1$ and the dynamics is renewed. The blue line describes a numerical simulation of the Langevin equation~\eqref{eq:srgbm-microscopic}, whereas the orange line is the solution~\eqref{eq:srgbm-solution-1-0} and~\eqref{eq:srgbm-solution-2-0}. In this example, we set $\mu = 0.05$, $\sigma^2 = 0.02$ and 
$r = 0.16$.   \label{fig:srgbm-solution}}
\end{figure}

\subsection{Method of simulation}
 The basic ingredient used to numerically simulate srGBM is to generate a trajectory using Eq.~\eqref{eq:srgbm-microscopic}. This is done \textit{\`a la} Langevin. Concretely, to obtain the distribution of the position of the particle at time $t$, we discretize the time $t=n\Delta t$, where $n$ is an integer. We initialize the position of the particle at $x(0)=1$, and then, at each step ($\tau=1,\dots,n$), the particle can either reset or it can evolve according to the laws of GBM. Thus,
\begin{enumerate}
    \item with probability $1-r\Delta t$ ($r$ is the rate of resetting), the particle undergoes GBM so that
\begin{align}
    x(\tau\Delta t) = x((\tau-1)\Delta t)+ x((\tau-1)\Delta t) \left[\mu + \sigma \sqrt{\Delta t} \eta (\tau \Delta t) \right],
\end{align}
where $\eta (\Delta t)$ is a Gaussian random variable with mean $0$ and variance $1$, and $\Delta t$ is the microscopic time step;
\item with complementary probability $r\Delta t$, resetting occurs such that
\begin{align}
    x(\tau \Delta t) &=x(0)=1.\label{ret-d}
\end{align}
\end{enumerate}
\noindent

In \fref{fig:srgbm-solution} we have compared \eref{eq:srgbm-solution-2-0} with the Langevin simulation to find an excellent match.

\section{Non-equilibrium properties of GBM under stochastic resetting}
\label{GPM}

In this section, we discuss non-equilibrium properties of GBM subjected to stochastic resetting. We first present exact results for the moments at all times. Next, we discuss the non-equilibrium steady state of GBM under stochastic resetting.

\subsection{Moments}
Moments of srGBM can be computed easily by applying \textit{the law of total expectation}. In practice, the $m$-th moment is obtained by raising \eref{eq:srgbm-solution-2-0} to the $m$-th power and then averaging with respect to the noise and $f(t_l|t)$ respectively
\begin{align}
&\left\langle x^m(t)\right\rangle = x_0^m e^{\left(m \mu + m(m-1)\frac{\sigma^{2}}{2}\right) t } \left\langle e^{ -\left(m\mu + m(m-1)\frac{\sigma^{2}}{2}\right)  t_l(t)} \right\rangle_{t_l} \nonumber \\ 
&= \frac{x_0^m}{m\mu  + m(m-1)\frac{\sigma^{2}}{2} - r}\left[\mu\, e^{\left(m\mu + m(m-1)\frac{\sigma^{2}}{2} - r\right)t} - r \right].
\label{eq:gbm-resetting-moments}
\end{align}
In general, three regimes for the evolution of the $m$-th moment can be identified based on the relation between the drift, noise amplitude and the resetting rate. First, when $r > r_m \equiv m\mu + m (m-1) \frac{\sigma^2}{2}$, the $m$-th moment 
converges to a limiting value $r/\left(r-r_m\right)$. At $r = r_m$, a sharp transition occurs, and the moment diverges linearly in time, i.e., $\langle x^m(t) \rangle \sim 1 + r t$. For $r < r_m$, this divergence becomes exponential. Table~\ref{tab:moments-behavior} summarizes the relationship between the parameters and the resulting behavior for the first two moments, i.e., the ensemble average and the second moment. The different limiting points of divergence for the moments can be seen as a hallmark multiplicative property of srGBM. For completeness we present a complementary renewal based derivation for the moments in~\aref{app:srgbm-moments}.

\begin{table}[t!]
 \caption{\textbf{Moments behavior in srGBM. \label{tab:moments-behavior}}}
\begin{tabular}{|c|c|c|c|}
\hline
\multirow{3}{*}{\textbf{Moment}}  & \multicolumn{3}{|c|}{\textbf{Limiting behavior}} \\
           & Exponential  & Linear  & \multirow{2}{*}{Convergence} \\
           & divergence & divergence &\\\hline
$\langle x(t) \rangle$      &     $r < \mu$        &  $r = \mu$        &  $r > \mu$                   \\
Eq.~\eqref{eq:gbm-resetting-first-moment} & ($\sim e^{(\mu-r)t}$) & ($\sim r t$) & ($\sim r / (r-\mu)$)  \\\hline
$\langle x^2(t) \rangle$   &     $r < 2\mu + \sigma^2$        &  $r = 2\mu + \sigma^2$         &  $r > 2\mu +\sigma^2$                 \\
Eq.~\eqref{eq:gbm-resetting-second-moment}& ($\sim e^{(2\mu+\sigma^2-r)t}$) & ($\sim r t$) & ($\sim r / (r-2\mu- \sigma^2)$)\\\hline
\end{tabular}
\end{table}

\subsection{Probability Density Function} 

The probability density function (PDF) of a reset-process satisfies the following renewal equation \cite{evans2020stochastic}
\begin{align}\label{eq:pdf-solution resetting}
  P_r(x,t|x_0) &= e^{-rt}P_{0}(x,t|x_0)+r\int_{0}^{t}e^{-ru}P_{0}(x,u|x_0)\,du,
\end{align}
where $P_{0}(x,t|x_0)$ is the PDF of the reset-free ($r=0$) process and in case of GBM reads~\cite{aitchison1957lognormal,stojkoski2020generalised}
\begin{align}
P_{0}(x,t|x_0)&= \frac{1}{x\sqrt{2\pi \sigma^2 t}}  \exp \left( \frac{-\left[\log (\frac{x}{x_0})-(\mu-\frac{\sigma^2}{2})t\right]^2}{2\sigma^2 t}  \right).
\end{align}
The steady state is then found by taking Laplace transform of Eq.~\eqref{eq:pdf-solution resetting}, i.e., $ P_r^{ss}(x|x_0)=\lim_{t \to \infty} P_r(x,t|x_0)=r \hat{P}_{0}(x,r|x_0)$, where $\hat{P}_{0}(x,s|x_0) \equiv \int_0^\infty e^{-st} P_{0}(x,t|x_0)\,dt$. Following this (see~\aref{app:srgbm-pdf}), we find that the stationary distribution has a power law whose right tail is given by
\begin{align}
    P_r^{ss}(x|x_0) \sim C(x_0) x^{-\alpha - 1} \quad \textrm{if} \quad x > x_0,
    \label{NESS-srGBM}
\end{align}
for some normalizing constant $C(x_0)$ that is dependent on the initial condition and a shape parameter 
\begin{align}
    \alpha &= \frac{-(\mu - \sigma^2/2) + \sqrt{(\mu-\sigma^2/2)^2 + 2r \sigma^2}}{\sigma^2}. 
\end{align}


The attained stationarity is not enough to render the model ergodic. In standard GBM, non-ergodicity arises due to the noise induced fluctuations which exhibit a net-negative effect on the time-averaged particle position, but do not affect the ensemble average. Therefore, in order to observe stationary-like behavior on the long run one must track the evolution of an infinite number of trajectories~\cite{he2008random}. Introducing stochastic resetting does not alter this phenomenon. This is because the long time average of a finite sample of trajectories (defined below) will be dominated by extremely rare non-reset trajectories. However, as will be shown below, resetting represents an additional source of randomness that not only increases the net-negative effect on the time-averaged position but also under certain circumstances may induce a similar effect on the ensemble average. As a result, we observe a variety of long-time regimes for the sample average due to resetting. We discuss these issues next.

\section{Ergodic properties}
\label{ergodic}

In srGBM, the non-ergodicity of the sample average is manifested in the same way as in GBM, that is, by the difference between the time-average and ensemble growth rate \cite{peters2013ergodicity}. This is captured by the following estimator of the growth rate of a sample of GBM trajectories
\begin{align}
    g_{est}(t,N) &\equiv  \frac{1}{t}\log \left(\langle x(t) \rangle_N \right),
    \label{eq:growth-rate}
\end{align}
where 
\bea
\langle x(t) \rangle_N = \frac{1}{N} \sum_{i=1}^N x_i(t)
\label{PEA}
\eea
is known as the finite sample average with the property $\lim_{N \to \infty} \langle x(t) \rangle_N = \langle x(t) \rangle$. Similar estimator was used to study ergodic properties in continuous time random walk \cite{he2008random} and anomalous diffusion in disordered materials \cite{akimoto2016universal}.

The ensemble growth rate $\langle g \rangle$ is found by fixing the period $t$ and taking the limit as the sample grows infinitely, i.e., 
\begin{align}
\langle g \rangle &= \lim_{N\to \infty}  g_{est}(t,N).
    \label{eq:ensemble-growth-rate}
\end{align}
On the other hand, the time-average growth rate $\bar{g}$ is found by fixing the sample size $N$ and letting time remove the stochasticity,
\begin{align}
\bar{g} &= \lim_{t\to \infty}  g_{est}(t,N).
    \label{eq:time-average-growth-rate}
\end{align}
The non-ergodicity of the process is manifested in the \textit{non-commutativity} of the two limits. Concretely, it can be shown that the ensemble average growth rate is $\langle g \rangle = g(t)$, which can be obtained by substituting Eq.~\eqref{eq:gbm-resetting-moments} with $m=1$ in Eq.~\eqref{eq:growth-rate}. On the other hand, we find that the time average growth rate is $\bar{g} = 0$. Let us first present a proof for the simplest case $N=1$, and afterwards generalize the results for arbitrary sample sizes.
We start by substituting the solution $x(t)$ from \eref{eq:srgbm-solution-2-0} into \eref{eq:growth-rate} to obtain (setting $x_0=1$)
\begin{align}
   g_{est}(t,N=1) &= \left(\mu - \frac{\sigma^2}{2}\right) \left(1 - \frac{t_{l}}{t}\right) + \frac{\sigma}{t} \left(W(t) - W(t_{l})\right),
   \label{eq:gbm-resetting-time-average-growth-rate}
\end{align}
from where it follows that (\aref{app:srgbm-growth-rate})
\begin{align}
   \left\langle g_{est}(t,N=1) \right\rangle &= \left(\mu - \frac{\sigma^2}{2}\right) \left(1 - \frac{\langle t_{l} \rangle}{t}\right),
   \label{eq:time-average-growth-rate-first-moment}
   \end{align}
and
\begin{align}
\text{Var} \left[ g_{est}(t,N=1) \right] &= \left(\mu - \frac{\sigma^2}{2}\right)^2 \frac{\text{Var}\left[ t_{l} \right]}{t^2} + \frac{\sigma^2}{t} \left( 1 - \frac{\langle t_{l} \rangle}{t} \right),
   \label{eq:time-average-growth-rate-second-moment}
\end{align}
where `Var' stands for variance.
These results hold for any resetting time density. In particular, for Poissonian resetting, 
$\left\langle g_{est}(t,N=1) \right\rangle=\text{Var} \left[ g_{est}(t,N=1) \right]=0$ in the limit $t \to \infty$. This essentially implies that the distribution of $g_{est}(t,N=1)$ must converge to a Dirac delta function asymptotically. In other words, as $t \to \infty$, the observed growth rate $\bar{g}$ will differ from $0$ with probability zero.

The proof for arbitrary $N$ is based on extreme value theory \cite{peters2013ergodicity,majumdar2020extreme}. In particular, we will show that $\bar{g}$ is bounded from above and below, and that these bounds coincide.
The upper bound can be shown by observing that for a fixed $t$ and sample size $N$
\begin{align}
    g_{est}(t,N) \leq \max_{i} \frac{1}{t} \log x_i(t) = \max_i g_{est}^{\left[ i \right]}(t,N = 1),
\end{align}
since the system size is finite. Taking the limit with respect to time, it follows that
\begin{align}
    \bar{g} \leq \max_{i} \lim_{t \to \infty} g_{est}^{\left[ i \right]}(t,N = 1) = 0. 
\end{align}
In a similar manner, for the lower bound we have
\begin{align}
    \bar{g} \geq \min_{i} \lim_{t \to \infty} g_{est}^{\left[ i \right]}(t,N = 1) = 0. 
\end{align}
Hence, the bounds for $\bar{g}$ saturate to a threshold which is zero implying $\bar{g}=0$ for any fixed sample size.

\begin{figure}[t!]
\includegraphics[width=8.6cm]{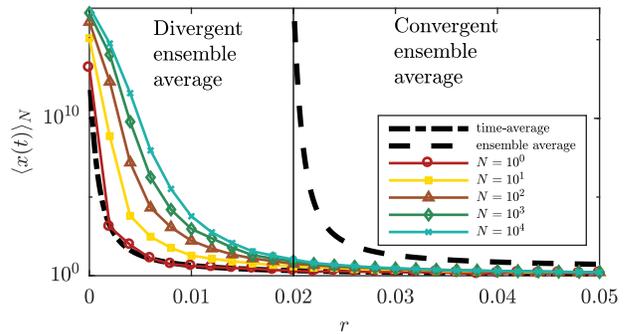} 
\caption{\textbf{Ergodicity breaking in srGBM}. Long time sample average ($t = 10^5$) as a function of $r$ for various sample sizes. For each sample size, $10^4$ random realizations were generated and the median results are shown. We set $\mu = 0.02$, $\sigma^2 = 0.01$. The black vertical line indicates the threshold $r=\mu$ where the ensemble average becomes convergent.
\label{fig:srgbm-ergodicity-breaking}}
\end{figure}

To numerically illustrate this non-ergodicity, we plot the long run sample average $\langle x(t) \rangle_N$ as a function of resetting rate $r$ for various sample sizes in Fig.~\ref{fig:srgbm-ergodicity-breaking}. We simulate  the sample average by generating $N$ independent and identical copies of the Langevin simulation, i.e.,
\bea
\langle x(\tau \Delta t) \rangle_N &= \frac{\sum_i^N x_i(\tau \Delta t)}{N}.
\eea
As described in the main text, $\langle x(\tau \Delta t) \rangle_N$ will resemble the ensemble average as long as $\tau \Delta t < t_c$, and afterwards it will collapse to its time-average behavior. 

For $r \leq \mu$, the ensemble average diverges (dashed black line). However, the
time-average is convergent resulting in the sample average to converge. Even in the regime when $r > \mu$ the sample average is closer to the time-average and there are apparent differences with the ensemble average. This is best seen in the single system (marked with a circle), which is dominated by the time-average behavior (dash-dotted black line). As the sample size increases, the sample average draws closer to the magnitude of the ensemble average but it always remains convergent.

To explain the differences in observations belonging to different sample sizes one can use an analogy with Random Energy Model (REM) studied by Derrida~\cite{derrida1981random,peters2018sum}. In REM, there exists a critical inverse temperature $t_c$ below which the quenched and annealed averages are identical whereas above $t_c$, only the quenched average is observed and the system is frozen in a small number of configurations of energy \cite{gueudre2014explore}. In srGBM, $t_c$ corresponds to a critical self-averaging time until which the sample average resembles the corresponding ensemble value, i.e., the time until Eq.~(\ref{eq:ensemble-growth-rate}) is valid. 
However, note that in the absence of resetting, the critical self-averaging time is strictly determined by and is proportional with the sample size. Hence, as the sample size increases the sample average will spend longer time resembling the ensemble average. These dynamics are accumulated and effectively reflected in the observed time-average at the end. In stark contrast, we show that in srGBM, $t_c$ depends on both the sample size and the resetting strength and thus resulting in different long-time regimes. This is discussed next.

\section{Self-averaging properties} 
\label{SAP}


In srGBM, the critical self-averaging time can be estimated by the relative variance of the sample average, namely,
\begin{align}
    \mathrm{R}_N(t) &\equiv \frac{\text{Var}(\langle x(t) \rangle_N)}{\langle \langle x(t) \rangle_N\rangle^2},
    \label{eq:relative-variance}
\end{align}
where $\langle \cdot \rangle$ and $\text{Var}(\cdot)$ notations, without $N$ as a subscript, refer to the averages over all possible sample average realizations. Using \eref{PEA} and the property for variance of sums of IID random variables, Eq.~\eqref{eq:relative-variance} can be rewritten as 
\begin{align}
   \mathrm{R}_N(t) &= \frac{1}{N} \frac{\langle x^2(t) \rangle - \langle x(t) \rangle^2}{\langle x(t) \rangle^2}.
   \label{eq:relative-variance-shortened}
\end{align}
If $\mathrm{R}_N(t) \ll 1$, the system is self-averaging, i.e., the sample average will be close to the ensemble average. Thus, the system will be self-averaging until the critical point $t_c$ which occurs at $\mathrm{R}_N(t_c) = 1$. We can use this information and rephrase \eref{eq:relative-variance-shortened} as
\begin{align}
    N+1 &= \frac{\langle x^2(t_c) \rangle}{\langle x(t_c) \rangle^2},
    \label{eq:critical-self-averaging-numerical}
\end{align}
which is the governing relation to determine $t_c$. This is done by plotting Eq.~\eqref{eq:critical-self-averaging-numerical} as a function of $r$ in Fig.~\ref{fig:gbm-resetting-critical-self-averaging}(a). Starting at $r=0$, the self-averaging time $t_c$ first decreases and then increases as a a function of resetting rate $r$. Depending on the trade-off between resetting rate ($r$), drift ($\mu$) and noise strength ($\sigma$), the system exhibits three different regimes which we explore in the following.

\subsection{Frozen state}
\begin{figure*}[ht!]
\includegraphics[width=17.4cm]{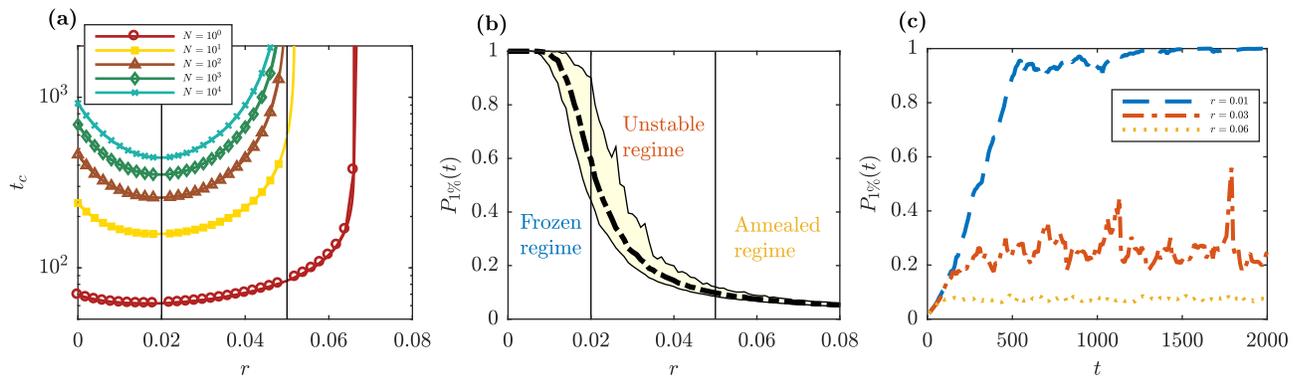}
\caption{\textbf{Self-averaging in srGBM}. \textbf{(a)} Critical self-averaging time $t_c$ as a function of $r$ for various sample sizes $N$. $t_c$ is seen to be minimized at $r^*=\mu$ for large sample size. The vertical lines indicate the thresholds $r=\mu$ and $r=2\mu+\sigma^2$ at which the ensemble average and second moment  respectively become convergent. \textbf{(b)} Long time ($t = 10^5$) probability of observing a microstate that is among the largest 1\% averaged across $10^4$ simulations. The median results are shown, and the filled region is the 5th and 95th percentile. Three regimes correspond to (i) $r<\mu$ (frozen), (ii) $2\mu +\sigma^2>r>\mu$ (unstable annealed) and (iii) $r>2\mu +\sigma^2$ (stable annealed). \textbf{(c)} Statistics of example simulations for the probability of observing a a microstate that is among the largest 1\%. Different colors indicate the regimes mentioned in (\textbf{b}). Parameters: $\mu = 0.02$, $\sigma^2 = 0.01$ and $N = 10^4$. \label{fig:gbm-resetting-critical-self-averaging}}
\end{figure*}

In the regime $r< \mu$, if we were to start with $N$-microstates with equally distributed energies, during the self-averaging period, inequality will increase and the system will eventually end up in a frozen configuration, as in REM. This is because both the ensemble average, given by \eref{eq:gbm-resetting-first-moment} [putting $m=1$ in \eref{eq:gbm-resetting-moments}], and the second moment, given by  \eref{eq:gbm-resetting-second-moment} [putting $m=2$ in \eref{eq:gbm-resetting-moments}], are divergent. Thus, we can respectively approximate them as 
\begin{align}
    \langle x(t) \rangle &\approx \frac{\mu}{\mu - r}\exp\left[\left(\mu - r\right)t\right] x_{0},
\end{align}
and
\begin{align}
     \langle x^2(t) \rangle \approx \frac{2\mu +\sigma^{2}}{2\mu + \sigma^{2}-r}\exp\left[(2\mu + \sigma^{2} - r)t\right] x_{0}^{2}.
     \label{eq:srgbm-msd-appendix}
\end{align}
Putting these two equations in~\eref{eq:critical-self-averaging-numerical} we can get an approximate equation for the critical self-averaging time in this regime as
\begin{align}
    t_c \approx \frac{1}{r+\sigma^2} \log \left[ (N+1)\frac{ \mu^2 (2 \mu + \sigma^2 -r)}{(\mu -r)^2(2 \mu + \sigma^2)}\right].
\end{align}
For a large enough sample this reduces to 
\begin{align}
    t_c \approx \frac{1}{r+\sigma^2} \log \left[ (N+1)\right],
\end{align}
which is precisely the behavior observed in Fig.~\ref{fig:gbm-resetting-critical-self-averaging}(a). 

We quantify the degree of freezing with the probability $P_{1\%}(t)$ that the system occupies a microstate that is among the largest 1\% of the sampled particle energies in time $t$ (Fig.~\ref{fig:gbm-resetting-critical-self-averaging}(b)). Numerically, this is easily done by relabeling the $N$ trajectories i.e., without loss of generality we assume that $x_1(\tau \Delta t) \geq x_2(\tau \Delta t) \geq \dots \geq x_N(\tau \Delta t) $. Then, $P_{1\%}(\tau \Delta t)$ in the period $\tau \Delta t$ is estimated as
\bea
P_{1\%} (\tau \Delta t) &= \frac{\sum_j^{N/100} x_j(\tau \Delta t)}{\sum_i^N x_i(\tau \Delta t)}.
\eea
In the economics literature, $P_{1\%}$ is interpreted as a measure of income inequality and its observed dynamics are expressed through changes in the model parameters, reflecting shocks (changes in model parameters due to external forces) in the system conditions \cite{bouchaud2000wealth,gabaix2016dynamics}. Thus, a value of $P_{1\%}$ closer to unity indicates a frozen configuration. An example for how $P_{1\%}$ behaves as a function of time is given in Fig.~\ref{fig:gbm-resetting-critical-self-averaging}(c) with a dashed line.

\subsection{Unstable state}
 In the regime when $\mu < r < 2\mu + \sigma^2$ the ensemble average, given by \eref{eq:gbm-resetting-first-moment}, is converges to a stationary value, whereas the second moment, given by \eref{eq:gbm-resetting-second-moment}, remains divergent. Specifically, then the ensemble average is approximated as
\begin{align}
    \langle x(t) \rangle &\approx \frac{r}{ r - \mu} x_{0},
    \label{eq:srgbm-convergent-value-1st-moment}
\end{align}
while the evolution of the second moment remains the same as in~\eref{eq:srgbm-msd-appendix}. Putting these two equations in~\eref{eq:critical-self-averaging-numerical}, we find
the critical self-averaging time to be
\begin{align}
    t_c \approx \frac{1}{2\mu+\sigma^2 - r} \log \left[ (N+1)\frac{ r^2 (2 \mu + \sigma^2 -r)}{(r - \mu)^2(2 \mu + \sigma^2)}\right].
\end{align}
Again, for a large enough sample the above equation reads
\begin{align}
    t_c \approx \frac{1}{2\mu+\sigma^2 - r} \log \left[ (N+1)\right].
\end{align}
which is clearly an increasing function of $r$, as can be seen in Fig.~\ref{fig:gbm-resetting-critical-self-averaging}(a). In words, beyond $r>\mu$, increments in $r$ reflect in an increased self-averaging time.
Since $t_c$ is a decreasing function of resetting rate $r$ when $r < \mu$ and an increasing function when $ \mu < r < 2\mu + \sigma^2$, it must be the case that the function attains a a local minimum on the interval $\left[ 0, 2\mu+\sigma^2 \right]$ at the point $r^* = \mu$.  Note that the transition point coincides with the threshold required for the ensemble average to become convergent. This is indeed observed in \fref{fig:gbm-resetting-critical-self-averaging}(a). We remark that for a small sample size, the optimal resetting rate $r^{*}$ that minimizes $t_c$, is weakly dependent on the sample size $N$. Therefore, optimization of $t_c$ with respect to resetting rate $r$ is one of the central features of this work. 

Moreover, in this state, the resetting rate is large enough to squeeze the effect of the drift and constrain the ensemble average, but the second moment remains divergent. Due to this, after the self-averaging period even if the particle's position is reset, it will quickly return to its pre-resetting position. Thus the system will not always be trapped in a small number of configurations (Fig.~\ref{fig:gbm-resetting-critical-self-averaging}(b)). Instead, the share $P_{1\%}$  will  randomly phase between large and small values over time. This, in turn implies that the state of the configurations will be unstable over time as evidenced in Fig.~\ref{fig:gbm-resetting-critical-self-averaging}(c), with the dash dotted line. This is a significant impact of resetting on the long time behavior of GBM. 

\subsection{Stable state}

As $r$ is further increased and above $2\mu + \sigma^2$, the second moment also becomes convergent and a third regime appears. In this case, in the long time limit the second moment is convergent and reads
\begin{align}
    \langle x^2(t) \rangle \approx \frac{r}{ r - 2\mu - \sigma^2} x_{0},
\end{align}
By combining the above equation with Eq.~\eqref{eq:srgbm-convergent-value-1st-moment}, we get that the long time relative variance in this regime is approximately constant
\begin{align}
    R_N(t) \approx \frac{1}{N} \left[ \frac{(r-\mu)^2}{r(r-2\mu-\sigma^2)} -1\right].
    \label{eq:srgbm-relative-variance-annealing-regime}
\end{align}
It can be observed that in this case the sample average resembles the ensemble value (Fig.~\ref{fig:gbm-resetting-critical-self-averaging}(a)) and exhibits diverse configurations (Fig.~\ref{fig:gbm-resetting-critical-self-averaging}(b)). Consequently, the probability $P_{1\%}$ of observing extreme configurations stabilizes over time (Fig.~\ref{fig:gbm-resetting-critical-self-averaging}(c) dotted line). This is the onset of a stable annealed regime. In this state, for a large enough sample size $N$, the system may forever mimic ergodic behavior.
More precisely, self-averaging will always occur in the system if $ R_N(t)$ in Eq.~\eqref{eq:srgbm-relative-variance-annealing-regime} is less than one. This leads to the condition
\begin{align}
    N > \frac{\mu^2+ r \sigma^2}{(r - 2\mu -\sigma^2) r},
\end{align}
 This is remarkably different from the previous two regimes, where for any fixed sample size, we would eventually observe discrepancies between the ensemble and sample averages, and is another important feature induced by resetting on GBM.

\section{Conclusion} 
\label{conclusions}

In this work, we performed a detailed analysis on the spatial and ergodic properties of srGBM. While discrete time stochastic and deterministic multiplicative processes with resetting have been studied in \cite{manrubia1999stochastic,zanette2020fat}, a detailed and systematic investigation for the spatial and ergodic properties for the continuous time multiplicative process such as GBM was still missing. The emergence of three regimes namely
frozen/quenched, an unstable and a stable annealed state is especially noteworthy. The ensemble properties of the second and third regime have been explored to a great extent in the income inequality literature~\cite{gabaix2016dynamics}. Indeed, most identified power laws in nature have exponents such that the average is well-defined but the variance is not, implying that the second regime is an expected outcome~\cite{newman2005power}.
Nonetheless, recent studies in economics also identify non-ergodic and divergent behavior in samples of srGBM trajectories thus suggesting the existence of the first regime~\cite{berman2020wealth,peters2019ergodicity}. A typical example of srGBM would be investment portfolios with reconstitution (addition or removal of constituents) \cite{peters2011optimal} where one might observe a such multi-stable landscape. Naturally, the results presented here for srGBM lend themselves as a baseline to depict the long run behavior of the above-mentioned scenarios.
This empirical investigation represents an intriguing research question which we leave for future work.

From a technical perspective, it is important to stress that the solution \eqref{eq:srgbm-solution-2-0}, time-average growth rate \eqref{eq:gbm-resetting-time-average-growth-rate}, and the critical self-averaging time \eqref{eq:critical-self-averaging-numerical} are universal and do not depend on the resetting time density. It remains to be seen how the statistical properties of GBM alter intricately under arbitrary resetting time density. Moreover, the dependence of $t_c$ on generic resetting time distribution suggests that we may observe diverse properties for the optimality based on the resetting strategy that we employ (similar to various optimization of the mean first passage time under resetting \cite{pal2017first}). Finally, GBM, besides being a canonical model for self-reproduction, is also used to describe diffusion processes where the particle spreads very fast, such as heterogeneous and turbulent diffusion~\cite{baskin2004superdiffusion,sandev2020hitting}. Exploring the applications of resetting strategies on GBM thus also represents a potential research avenue that is of broad interest.

\section*{Acknowledgements}
VS, TS and LK acknowledge financial support by the German Science Foundation (DFG, Grant number ME 1535/12-1). TS was supported by the Alexander von Humboldt Foundation. AP gratefully acknowledges support from the Raymond and Beverly Sackler Post-Doctoral Scholarship and the Ratner Center for Single Molecule Science at TelAviv University.

\appendix
\onecolumngrid

\section{Calculation of moments for srGBM}\label{app:srgbm-moments}

In the main text we showed how to derive the moments of srGBM using the law of total expectation. In this section, we present alternative derivations for the moments using Fokker-Planck and a renewal approach respectively. 

\subsection{Fokker-Planck approach}

The Fokker-Planck equation for the GBM with exponential resetting to the initial position $P_r(x,t=0|x_0)=\delta(x-x_0)$ reads 
\begin{align}\label{eq:GBM-resetting-Fokker_Planck}
\frac{\partial}{\partial t}P_r(x,t|x_0)=-\mu\frac{\partial}{\partial x}xP_r(x,t|x_0)+\frac{\sigma^2}{2}\frac{\partial^2}{\partial x^2}x^{2}P_r(x,t|x_0)-r P_r(x,t|x_0)+r P_r(x,t=0|x_0),
\end{align}
where $r$ is the rate of resetting to the initial position $x_0$. The last two terms in the RHS of \eref{eq:GBM-resetting-Fokker_Planck} represent respectively the loss of the probability from position $x\neq x_0$ due to the resetting and consecutively a gain in the probability at the initial position $x_0$ from all the other positions in space.
Taking Laplace transform on both sides of \eref{eq:GBM-resetting-Fokker_Planck} gives
\begin{align}\label{generalizedFPE_standard reset laplace}
s\hat{P_r}(x,s|x_0)-P_r(x,t=0|x_0)=-\mu\frac{\partial}{\partial x}x\hat{P_r}(x,s|x_0)+\frac{\sigma^2}{2}\frac{\partial^2}{\partial x^2}x^{2} \hat{P_r}(x,s|x_0)-r \hat{P_r}(x,s|x_0)+\frac{r}{s}P_r(x,t=0|x_0),
\end{align}
where $\hat{g}(s)=\mathcal{L}[g(t)]=\int_{0}^{\infty}g(t)e^{-st}\,dt$ is the Laplace transform of $g(t)$.
Rewriting the above equation, one gets
\begin{align}\label{eq:generalizedFPE_standard reset laplace2}
s\hat{P_r}(x,s|x_0)-P_r(x,t=0|x_0)=s\times\frac{1}{s+r}\left[-\mu\frac{\partial}{\partial x}x\hat{P_r}(x,s|x_0)+\frac{\sigma^2}{2}\frac{\partial^2}{\partial x^2}x^{2}\hat{P_r}(x,s|x_0)\right],
\end{align}
which, upon an inverse Laplace transform, gives us the following convoluted equation
\begin{align}\label{generalizedFPE reset}
\frac{\partial}{\partial t}P_r(x,t|x_0)=\frac{\partial}{\partial t}\int_{0}^{t}\eta(t-t')&\left[-\mu\frac{\partial}{\partial x}x P_r(x,t'|x_0)+\frac{\sigma^2}{2}\frac{\partial^2}{\partial x^2}x^{2}P_r(x,t'|x_0)\right]dt',
\end{align}
where $\eta(t)=e^{-r t}$. Note that a similar equation was also used in Ref.~\cite{stojkoski2020generalised} to explore the properties of a generalized GBM process subject to subdiffusion (without resetting). In what follows, we would like to write a dynamical equation for the moments $\langle x^{m}(t)\rangle \equiv \int_0^\infty~x^m(t)P_r(x,t|x_0)~dx$ using \eref{generalizedFPE reset}.
To see this, we multiply both sides of Eq.~(\ref{generalizedFPE reset}) by $x^m$ and integrate over $x$ to find 
\begin{align}\label{m-th moment}
\frac{\partial}{\partial t}\langle x^{m}(t)\rangle=\left[\frac{\sigma^{2}}{2}m(m-1)+\mu\,m\right]\mu\,\frac{d}{dt}\int_{0}^{t}\eta(t-t')\langle x^{m}(t')\rangle\,dt'.
\end{align}
In Laplace space, the solution to this equation reads
\begin{align}\label{m-th moment laplace app}
\langle \hat{x}^{m}(s)\rangle=\frac{s^{-1}}{1-\hat{\eta}(s)\left[\frac{\sigma^{2}}{2}m(m-1)+\mu\,m\right]}x_{0}^{n},
\end{align}
where $\hat{\eta}(s)=\mathcal{L}\left[e^{-rt}\right]=\frac{1}{s+r}$, and $x_0=x(0)$ is the fixed initial condition. For $m=1$, 
the solution of the equation for the ensemble average (first moment or the mean value) in Laplace space is given by
\begin{align}\label{mean general eta app}
\left\langle \hat{x}(s)\right\rangle=\frac{s^{-1}}{1-\mu\hat{\eta}(s)} x_{0}=\frac{s^{-1}}{1-\mu/(s+r)}x_{0},
\end{align}
which can be inverted to obtain the following expression for the mean
\begin{align}\label{eq:gbm-resetting-first-moment}
\left\langle x(t)\right\rangle 
=\frac{x_0}{\mu - r}\left[\mu\, e^{\left(\mu - r\right)t} - r \right],
\end{align}
Similarly, for $m=2$, using \eref{m-th moment}, we obtain the following equation for the second moment
\begin{align}
\frac{\partial}{\partial t}\langle x^2(t)\rangle=(\sigma^{2}+2\mu)\frac{d}{dt}\int_{0}^{t}\eta(t-t')\langle x^2(t')\rangle\,dt',
\end{align}
with a solution in Laplace space
\begin{align}
\langle \hat{x}^2(s)\rangle=\frac{s^{-1}}{1-(\sigma^{2}+2\mu)\hat{\eta}(s)} x_0^2=\frac{s^{-1}}{1-(\sigma^{2}+2\mu)/(s+r)}x_0^2,
\end{align}
which can be inverted to obtain the following expression for the second moment
\begin{align}\label{eq:gbm-resetting-second-moment}
\left\langle x^{2}(t)\right\rangle 
=\frac{ x_{0}^{2}}{2\mu + \sigma^{2}-r}\left[(2\mu +\sigma^{2})\, e^{(2\mu + \sigma^{2} - r)t} -r \right].
\end{align} 
Finally, inverting \eref{m-th moment laplace app}, we arrive at \eref{eq:gbm-resetting-moments} as was mentioned in the main text.

\subsection{Renewal approach}
It is now well understood that resetting is a renewal process in the sense the process erases its memory after each resetting. This leads to an advantage since the solution of the reset-process $P_r(x,t|x_0,0)$ can be written in terms of the underlying reset-free process $P_0(x,t|x_0,0)$. Following Ref.~\cite{evans2020stochastic}, we can write
\begin{align}
\label{renewal-SM-1}
P_r(x,t|x_0,0)&=e^{-rt}P_{0}(x,t|x_0,0)+\int_0^t~re^{-ru}~P_{0}(x,u|x_0,0)\,du,
\end{align}
which is essentially Eq.~\eqref{eq:pdf-solution resetting}. This equation can be interpreted in terms of a renewal process, i.e., each  resetting  event  to  the  initial  position $x_0$ renews the process at a rate $r$. Between two consecutive renewal events, the particle undergoes its original dynamics. In fact \eref{renewal-SM-1} can also be obtained from \eref{generalizedFPE reset} using a subordination approach used in \cite{stojkoski2020generalised}. To show this, 
note that the solution of Eq.~(\ref{generalizedFPE reset}) can be represented by the subordination integral
\begin{align}
    P_r(x,t|x_0)=\int_{0}^{\infty}P_{0}(x,u|x_0)\,h(u,t)\,du,
\end{align}
where $h(u,t)$ is a subordination function. 
For the present case of exponential waiting time for resetting, the subordination function for srGBM reads, see Ref.~\cite{stojkoski2020generalised},
\begin{align}
\hat{h}(u,s)=\frac{1}{s\hat{\eta}(s)}e^{-\frac{u}{\hat{\eta}(s)}}=\frac{s+r}{s}e^{-u(s+r)} \quad \rightarrow \quad h(u,t)=e^{-rt}\delta(t-u)+r\,e^{-ru}\theta(t-u).
\end{align}
Thus
\begin{align}
    P_r(x,t|x_0) =& \int_{0}^{\infty} P_{0}(x,t|x_0,0)\left[e^{-rt}\delta(t-u)+r\,e^{-ru}\theta(t-u)\right]du
    \label{eq:log-normal-distribution}
\end{align}
from where we recover the renewal form given in~\eref{renewal-SM-1}.
Taking Laplace transform on the both sides of the above equation gives
\begin{align}\label{solution resetting laplace}
    \hat{P_r}(x,s|x_0)=\hat{P}_{0}(x,s+r|x_0)+\frac{r}{s}\hat{P}_{0}(x,s+r|x_0)=\frac{s+r}{s}\hat{P}_{0}(x,s+r|x_0),
\end{align}
where $\hat{P}_{0}(x,s)$ is the Laplace transform of the
underlying propagator. By multiplying Eq.~\eqref{solution resetting laplace} with $x^m(s)$ and integrating out $x$, the $m$-th moment of srGBM in Laplace space can be written as
\bea
 \left \langle \hat{x}^m(s) \right \rangle &=\frac{s+r}{s} 
 \left \langle \hat{x}^m(s+r) \right \rangle_{r=0},
 \label{moments-SM-renewal}
\eea
where $  \left \langle \hat{x}^m(s) \right \rangle_{r=0}$ is the $m$-th moment without resetting in the Laplace space. Note that the equations derived so far do not depend on the specific choice of underlying dynamics. Moving forward, we will turn our focus to the GBM process. In particular, the GBM-propagator reads
\begin{align}
P_{0}(x,t|x_0,0)&=\frac{1}{x\sqrt{2\pi \sigma^2 t}} \exp \left( -\frac{\left[\log (\frac{x}{x_0})-(\mu-\frac{\sigma^2}{2})t\right]^2}{2\sigma^2 t}  \right),
\label{renewal-1-log}
\end{align}
which is a log-normal distribution. The moments, obtained from \eref{renewal-1-log}, read
\bea
\left \langle x^m(t) \right \rangle_{r=0}=x_0^{m}\,e^{\left(\sigma^2m(m-1)/2+\mu m\right)t}.
\eea
Computing the Laplace transforms $\left \langle \hat{x}^m(s) \right \rangle_{r=0}$ from above and substituting
into \eref{moments-SM-renewal} gives us the moments of srGBM in Laplace space. Inverting them, we recover the results as given by Eq.~(\ref{eq:gbm-resetting-moments}).

\section{Full expression for the stationary distribution}\label{app:srgbm-pdf}

In this section, we provide the full expressions for the steady state. To this end, we recall \eref{renewal-SM-1} and take the limit $t \to \infty$. The first term on the RHS of \eref{renewal-SM-1} drops out and we are left with
\begin{align}
    P_r^{ss}(x|x_0)=\lim_{t\rightarrow\infty}P_r(x,t|x_0)=\int_0^\infty~re^{-ru}~P_{0}(x,u|x_0,0)\,du=r\hat{P}_{0}(x,r|x_0).
    \label{SS-srGBM-SM}
\end{align}
Thus to compute the steady state, we need the Laplace transform of the underlying propagator. In particular, for GBM, they can be computed from \eref{renewal-1-log}. Eventually, we have
\begin{align}\label{log normal pdf laplace}
    \hat{P}_{0}(x,s|x_0)=\frac{1}{\sqrt{(\mu-\sigma^{2}/2)^{2}+2\sigma^{2}s}}\left\lbrace
    \begin{array}{l l l}
     \smallskip & \left(\frac{x}{x_0} \right)^{-\frac{\sqrt{(\mu-\sigma^{2}/2)^{2}+2\sigma^{2}s}-(\mu-\sigma^{2}/2)}{\sigma^2} -1}, \quad & x>x_0, \\ 
     & \left(\frac{x}{x_0} \right)^{\frac{\sqrt{(\mu-\sigma^{2}/2)^{2}+2\sigma^{2}s}+(\mu-\sigma^{2}/2)}{\sigma^2} -1}, \quad & x\leq x_0.
\end{array}\right.
\end{align}
Thus, using \eref{SS-srGBM-SM}, we arrive at the following expressions for the non-equilibrium steady state for srGBM
\begin{align}\label{solution resetting long time}
     P_r^{ss}(x|x_0) =
     \frac{r \sigma^2}{\alpha \sigma^2 + \left(\mu - \frac{\sigma^2}{2}\right)}\left\lbrace\begin{array}{l l l}
     \smallskip & \left( \frac{x}{x_0}\right)^{-\alpha-1}, \quad & x>x_0, \\ 
     & \left( \frac{x}{x_0}\right)^{\alpha+2\left(\mu - \frac{\sigma^2}{2}\right)-1}, \quad & x\leq x_0,
\end{array}\right.
     \end{align}
where
\begin{align}
    \alpha &= \frac{-(\mu - \sigma^2/2) + \sqrt{(\mu-\sigma^2/2)^2 + 2r \sigma^2}}{\sigma^2},
\end{align}
is the shape parameter.
The right tail ($x>x_0$) of this result has been highlighted in \eref{NESS-srGBM} in the main text.


\section{Calculation of moments for the srGBM growth rate \label{app:srgbm-growth-rate}}

Here we derive the moments of the srGBM growth rate given in Eqs.~\eqref{eq:time-average-growth-rate-first-moment} and~\eqref{eq:time-average-growth-rate-second-moment}. We start by noting that the estimator of the growth rate when $N=1$ can be found by inputting Eq.~\eqref{eq:srgbm-solution-2-0} in Eq.~\eqref{eq:growth-rate}, i.e.,
\begin{align}
   g_{est}(t,N=1) &= \left(\mu - \frac{\sigma^2}{2}\right) \left(1 - \frac{t_{l}}{t}\right) + \frac{\sigma}{t} \left(W(t) - W(t_{l})\right),
   \label{eq:srgbm-growth-rate}
\end{align}
where for simplicity we have set $x(0) = 1$. In order to derive the moments of srGBM, we are going to utilize three basic properties of the Wiener process. That is, the process is characterized with a first moment, $\langle W(t) \rangle = 0$, second moment $\langle W^2(t) \rangle = t$, and a covariance $\langle W(t) W(s) \rangle = \min \{ t,s\}$. 

Using this information, we can average Eq.~\eqref{eq:srgbm-growth-rate} first with respect to the Wiener noise, and then with respect to $t_l$ to get the first moment
\begin{align}
      \left\langle g_{est}(t,N=1) \right\rangle &= \left(\mu - \frac{\sigma^2}{2}\right) \left(1 - \frac{\langle t_{l}\rangle}{t}\right).
\end{align}

To derive the variance, we first square \eref{eq:srgbm-growth-rate}, and get
\begin{align}
    g_{est}^2(t,N=1) &=    \left(\mu - \frac{\sigma^2}{2}\right)^2 \left(1 - \frac{t_{l}}{t}\right)^2 + 2 \left(\mu - \frac{\sigma^2}{2}\right) \left(1 - \frac{t_{l}}{t}\right) \frac{\sigma}{t} \left(W(t) - W(t_{l})\right) + \frac{\sigma^2}{t^2} \left(W(t) - W(t_{l})\right)^2.
    \label{sec-mom-rand}
\end{align}
Again, we take the average of Eq.~\eqref{sec-mom-rand} first with respect to the Wiener noise, and then with respect to $t_l$. The computation goes as follows
\begin{align}
  \left\langle  g_{est}^2(t,N=1) \right\rangle &=    \left(\mu - \frac{\sigma^2}{2}\right)^2 \left\langle \left(1 - \frac{t_{l}}{t}\right)^2 \right\rangle + 2 \left(\mu - \frac{\sigma^2}{2}\right) \left(1 - \frac{\langle t_{l}\rangle }{t}\right) \frac{\sigma}{t} \left(\left\langle W(t)\right \rangle - \left\langle W(t_{l})\right \rangle\right) + \frac{\sigma^2}{t^2} \left\langle \left(W(t) - W(t_{l})\right)^2\right \rangle \\
  &=  \left(\mu - \frac{\sigma^2}{2}\right)^2 \left\langle \left(1 - \frac{t_{l}}{t}\right)^2 \right\rangle + \frac{\sigma^2}{t^2}  \left(\left\langle W^2(t)\right \rangle - \left \langle W(t)W(t_{l}) \right \rangle  + \left \langle W^2(t_{l} \right \rangle\right) \\
    &=  \left(\mu - \frac{\sigma^2}{2}\right)^2 \left\langle \left(1 - \frac{t_{l}}{t}\right)^2 \right\rangle + \frac{\sigma^2}{t}  \left( 1 - \frac{\langle t_l \rangle}{t}\right).
\end{align}
where we have used properties of the Wiener process.
Finally, variance of $g_{est}(t,N=1)$, as given in Eq.~\eqref{eq:time-average-growth-rate-second-moment} in the main text, is recovered from the following
\begin{align}
    \text{Var} \left[ g_{est}(t,N=1) \right] &=   \left\langle  g_{est}^2(t,N=1) \right\rangle -   \left\langle g_{est}(t,N=1) \right\rangle^2 \\
    &= \left(\mu - \frac{\sigma^2}{2}\right)^2 \frac{\text{Var}\left[ t_{l} \right]}{t^2} + \frac{\sigma^2}{t} \left( 1 - \frac{\langle t_{l} \rangle}{t} \right),
\end{align}
where $\text{Var}\left[ t_{l} \right]=\langle t_l^2 \rangle-\langle t_l \rangle^2$. 

\twocolumngrid

\providecommand{\noopsort}[1]{}\providecommand{\singleletter}[1]{#1}%

\end{document}